\documentclass[12pt]{iopart}
\usepackage{graphicx}
\usepackage{booktabs}
\usepackage{upgreek}
\usepackage{cite}
\usepackage{hyperref}

\usepackage[symbol]{footmisc}
\DefineFNsymbols*{arabic}{{$^1$}{$^2$}{$^3$}{$^4$}{$^5$}{$^6$}{$^7$}{$^8$}{$^9$}{$^{10}$}{$^{11}$}{$^{12}$}{$^{13}$}{$^{14}$}{$^{15}$}{$^{16}$}{$^{17}$}{$^{18}$}{$^{19}$}{$^{20}$}}
\setfnsymbol{arabic}

\newcommand{\ket}[1]{\left| #1\right\rangle}

\newcommand{\tha}{Trojan-horse attack }

\begin{document}

\title{Trojan-horse attacks threaten the security of practical quantum cryptography}

\author{Nitin Jain$^{1,3}$, Elena Anisimova$^{2}$, Imran Khan$^{1,3}$, Vadim Makarov$^{2}$, Christoph Marquardt$^{1,3}$, and Gerd Leuchs$^{1,3}$}

\address{$^1$ {Max Planck Institute for the Science of Light, G\"unther-Scharowsky-Str. 1/Bau 24, 91058 Erlangen, Germany}}
\address{$^2$ {Institute for Quantum Computing, University of Waterloo, 200 University Avenue West, Waterloo, Ontario N2L 3G1, Canada}}
\address{$^3$ {Friedrich-Alexander University Erlangen-N\"urnberg (FAU), Institute for Optics, Information and Photonics, Staudtstrasse 7/B2, 91058 Erlangen, Germany}}
\date{\today}

\begin{abstract}
A quantum key distribution system may be probed by an eavesdropper Eve by sending in bright light from the quantum channel and analyzing the back-reflections. We propose and experimentally demonstrate a setup for mounting such a Trojan-horse attack. We show it in operation against the quantum cryptosystem Clavis2 from ID~Quantique, as a proof-of-principle. With just a few back-reflected photons, Eve discerns Bob's secret basis choice, and thus the raw key bit in the Scarani-Ac\'in-Ribordy-Gisin 2004 protocol, with higher than $90\%$ probability. This would clearly breach the security of the cryptosystem. Unfortunately in Clavis2 Eve's bright pulses have a side effect of causing high level of afterpulsing in Bob's single-photon detectors, resulting in a high quantum bit error rate that effectively protects this system from our attack. However, in a Clavis2-like system equipped with detectors with less-noisy but realistic characteristics, an attack strategy with positive leakage of the key would exist. We confirm this by a numerical simulation. Both the eavesdropping setup and strategy can be generalized to attack most of the current QKD systems, especially if they lack proper safeguards. We also propose countermeasures to prevent such attacks.
\end{abstract}



\setcounter{footnote}{0}

\section{Introduction}
Quantum key distribution (QKD) provides a method to solve the task of securely distributing symmetric keys between two parties Alice and Bob~\cite{BB84,Gisin2002,Scarani2009a}. The security of QKD is based on the principles of quantum mechanics: an adversary Eve attempting to eavesdrop on the quantum key exchange inevitably introduces errors that warn Alice and Bob about her presence. In the last decade however, several vulnerabilities and loopholes in the physical implementations of QKD have been discovered, and proof-of-principle attacks exploiting them have shown the possibilities that Eve may get hold of the secret key without alerting Alice and Bob~\cite{Vakhitov2001,Gisin2006,Nauerth2009,Lydersen2010,Li2011,Wiechers2011,Jain2011,Jiang2012}. 

In most cases, vulnerabilities and loopholes arise due to technical imperfections or deficiencies of the hardware. For instance, no optical component can \emph{perfectly} transmit, or \emph{completely} absorb light. An optical pulse launched into a network of optic and optoelectronic components, e.g.,\ a QKD system, encounters several sites of Fresnel reflection and Rayleigh scattering~\cite{Saleh2007}. Some light thereby travels opposite to the propagation direction of the input optical signal. The properties and functionality of some component inside a QKD system may thus be probed from the quantum channel by sending in sufficiently-bright light and analyzing the back-reflected light. This forms the basis of a Trojan-horse attack~\cite{Bethune2000,Vakhitov2001,Gisin2006}.

Neither the concept, nor the danger of a Trojan-horse attack on QKD systems is new \cite{Bethune2000,Vakhitov2001,Gisin2006}. Also, it is the Alice device that is typically considered vulnerable to this kind of attacks since it prepares the quantum state in most QKD schemes. If a QKD system is operating, e.g.,\ the Bennett-Brassard 1984 (BB84) protocol~\cite{BB84}, then by sending a suitably-prepared bright pulse inside Alice and analyzing its back-reflections, Eve could obtain information about the setting of the polarizer~\cite{Bennett1992,Breguet1994a,Townsend1998} or the phase modulator~\cite{Rarity1992a,Muller1997,Bethune2000} responsible for encoding the secret bit. 

A simple way to detect a \tha red-handed is to install a passive monitoring device at Alice's entrance. This is usually implemented by a suitable detector (or an array of detectors) that measures different parameters of an incoming signal and raises an alarm whenever certain pre-characterized thresholds are crossed. However, a similar countermeasure cannot be straightforwardly adopted for the Bob device since it typically detects the \textit{already-quite-weak} states of light coming from the quantum channel -- a passive monitoring device would introduce unwanted attenuation and bring the secret key rates down further. Another countermeasure~\cite{Vakhitov2001,Gisin2006,Walenta2013,qkdstdzn} is to add an optical isolator to block the bright Trojan pulse from entering; however, this is not applicable to two-way systems such as plug-and-play schemes~\cite{Stucki2002a}. 

For the BB84 protocol, this does not pose a problem as Bob publicly declares his basis choice, i.e.,\ the setting of his polarizer/phase modulator. However, in the Scarani-Ac\'in-Ribordy-Gisin 2004 (SARG04) protocol~\cite{Scarani2004a,Branciard2005}, the secret bit is given by Bob's basis choice. If Eve can surreptitiously read Bob's phase modulator setting (= $0$ or $\pi/2$) from the quantum channel via a Trojan-horse attack, then she acquires knowledge of the raw key~\cite{Makarov2006}. She can then apply the same operations (sifting, error correction and privacy amplification~\cite{Bennett1992,Gisin2002,Scarani2009a}) as Alice and Bob and therefore, eavesdrop without being discovered and hence break the security of the system. 

SARG04 is more robust than BB84 against photon-number-splitting attacks~\cite{Brassard2000,Jiang2012}, which is useful for QKD systems such as Clavis2~\cite{clavis2guide} that employ attenuated laser sources. In the following sections however, we show that it can be vulnerable to Trojan-horse attacks on Bob. We believe this is the first proof-of-principle demonstration of such an attack on a practical QKD system (although static phase readout in Alice has been demonstrated before~\cite{Vakhitov2001,Gisin2006}, the previous experiments were not real-time and did not analyse the complete system). Furthermore, both our eavesdropping setup and strategy are universal: with simple modifications, they could be applied against entanglement-based, continuous-variable, or even the very recent measurement-device-independent QKD systems~\cite{Jouguet2013,Khan2013a,Liu2013,Silva2013} if they lack proper safeguards against Trojan-horse attacks. In such cases, it may be used even to break the BB84 protocol. 
\section{Theory and preparatory measurements}\label{sec:thry}
To prepare for a practical Trojan-horse attack, the eavesdropper Eve needs to know the answers to (at least) the following questions: 
\begin{enumerate}
\item What time should a Trojan-horse pulse be launched by Eve into Bob? 
\item What time would a back-reflected pulse of interest exit Bob and arrive on the quantum channel? And with what amplitude? 
\item What properties may be analyzed in a back-reflected pulse? 
\item How to avoid being detected by Alice and Bob? 
\item What is the most suitable wavelength for attack? 
\end{enumerate}
These questions are closely interrelated, and the answers to them naturally depend on the QKD system under attack. In this section, we address them specifically for Clavis2, the plug-and-play QKD system from ID~Quantique; or to be more precise, with the aim of crafting and executing an attack on Clavis2-Bob while it runs SARG04. 
\begin{figure}
\centering
\includegraphics[scale=0.39]{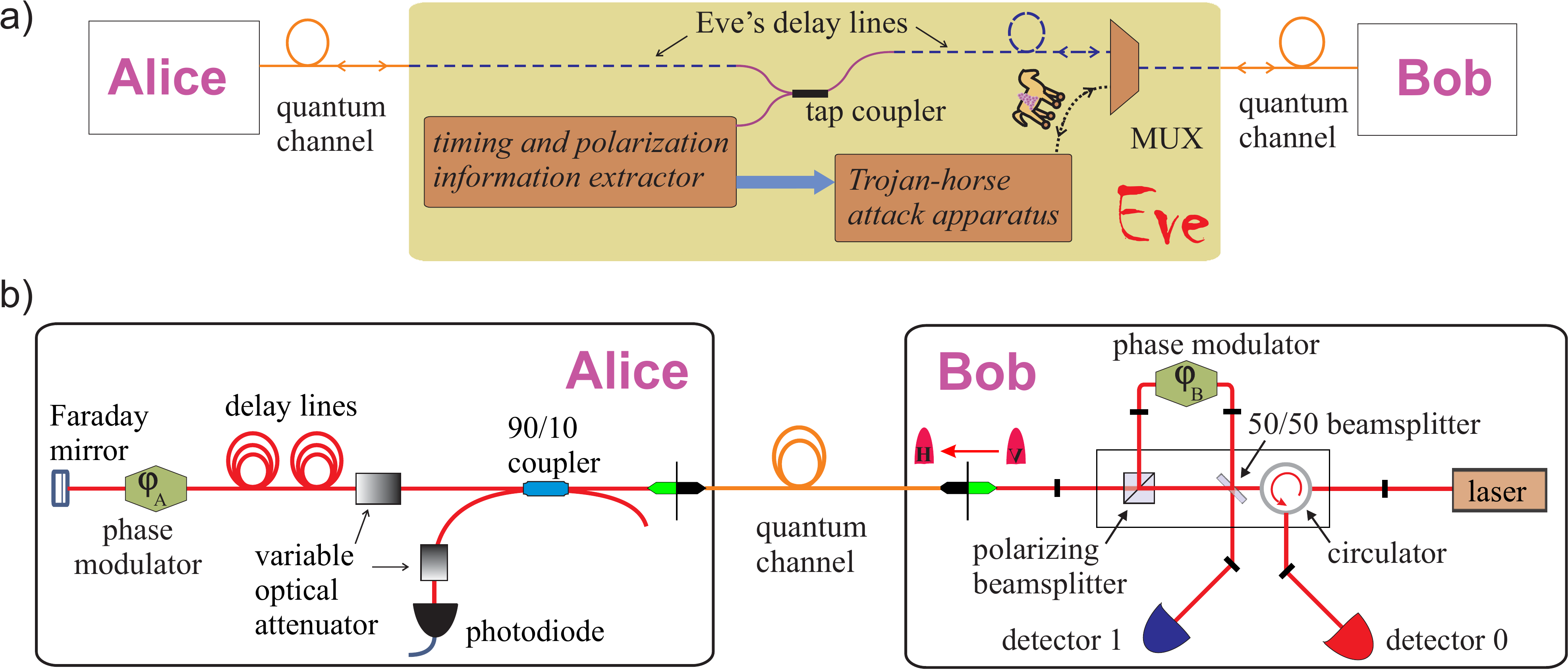}
\caption{Basic optical schematic of the Trojan-horse attack and plug-and-play QKD system. (a) Using the MUX, Eve multiplexes (in time and wavelength) the Trojan-horse pulses to the quantum signals traveling from Alice to Bob for probing Bob's basis choice. Reflections from Bob travel back to the Trojan-horse attack apparatus after being demultiplexed at the MUX. Eve may also replace parts of the quantum channel (in solid-orange) with her own delay lines (in dashed-blue). (b) A folded Mach-Zehnder interferometer operating in double pass facilitates a passive autocompensation of optical fluctuations (arising in the quantum channel) and forms the essence of plug-and-play schemes. Bob contains both the laser and single-photon detectors connected to his \textit{local} interferometer by means of a polarizing beamsplitter, 50/50 beamsplitter, and circulator (henceforth referred to as the PBS-BS-C assembly). Alice employs a Faraday mirror to reflect back the signals sent by Bob. The small black rectangles in Bob denote a pair of FC/PC connectors inside a mating sleeve.} 
\label{fpnp}
\end{figure}
Figure~\ref{fpnp}(a) shows the basic scheme of the attack while figure~\ref{fpnp}(b) shows the optical schematic of Clavis2 that operates in a two-way configuration based on the plug-and-play principle~\cite{Stucki2002a}. We briefly describe the principle below, and in the appendix we discuss several (technical) details via a numerical simulation. 

Bob contains both the laser and the detectors; he sends bright pulse pairs to Alice who prepares the quantum states and sends them back to Bob. For this, she randomly modulates the relative phase $\varphi_A = \{0,\pi/2,\pi,3\pi/2\}$ between the optical modes of each pair, and applies an attenuation so that the mean photon number of the resultant weak coherent pulses (returning to Bob on the quantum channel) is as dictated by the protocol. For SARG04, the optimal value is $\mu_{\rm SARG04} = 2\sqrt{T}$, where $T$ is the channel transmission~\cite{Branciard2005}. Bob applies a binary modulation chosen randomly per pair ($\varphi_B = 0$ or $\pi/2$, corresponding to the secret bit $\texttt{0}_b$ or $\texttt{1}_b$ respectively) and his pre-calibrated~\cite{Jain2011} gated detectors measure Alice's quantum states. The actual transmission uses the concept of \textit{frames}, a train of pulses that entirely fit in Alice's delay line in order to prevent errors that would otherwise result from Rayleigh backscattering~\cite{Stucki2002a}. A frame in our Clavis2 system is configured to be $215\, \upmu$s long, while the inter-frame separation depends on the total distance between Alice and Bob\footnote{Lower bound is provided by the delay line in Alice, which for our system results in $\sim 235\, \upmu$s.}.
\subsection*{Time of launching the Trojan-horse pulse}\label{tsndTHP}
Eve launches a Trojan-horse pulse (THP) into Bob at time $t_{E\rightarrow B}$ chosen so that the onward pulse and/or one of its back-reflections (from some component or interface inside Bob) travel through Bob's phase modulator (PM) while he is applying a voltage on it. As will be explained below, the back-reflected pulse coming out from Bob onto the quantum channel then carries an imprint of whatever random phase shift $\varphi_B$ had been applied by Bob. The time $t_{E\rightarrow B}$ is of course relative to events inside Bob repeating at $\textit{f}_B = 5\,$MHz. 
To be synchronized to the clock in Bob, Eve may steal a few photons from the bright pulses traveling to Alice using a tap coupler, as shown in figure~\ref{fpnp}(a). She can extract information such as timing and polarization from the measurement of these photons and use it in the preparation of the THPs. 
\subsection*{Time of arrival and amplitude of the back-reflected pulse}
\noindent As illustrated in figure~\ref{fpnp}(b), Bob comprises of a miscellany of fiber-optical components. This offers several interfaces from where (measurable) back-reflections could arise. Also, due to the asymmetric interferometer, there may be two different paths traversable in either directions, i.e.,\ for the arrival of the Trojan-horse pulse into Bob, and departure of a given reflection to the quantum channel. In essence, for a single THP sent into Bob, multiple reflections varying in time and amplitude can be expected. By means of repetitive measurements, a reflection-map for Bob -- temporal distribution of the back-reflection levels -- can be constructed. This is a task perhaps best suited for an optical time domain reflectometry (OTDR) device~\cite{Beller1998}. We obtained OTDR traces, or reflection-maps for Bob, for three different wavelengths: 806 nm, 1310 nm and 1550 nm. 
\begin{figure}
\centering
\includegraphics[scale=0.5]{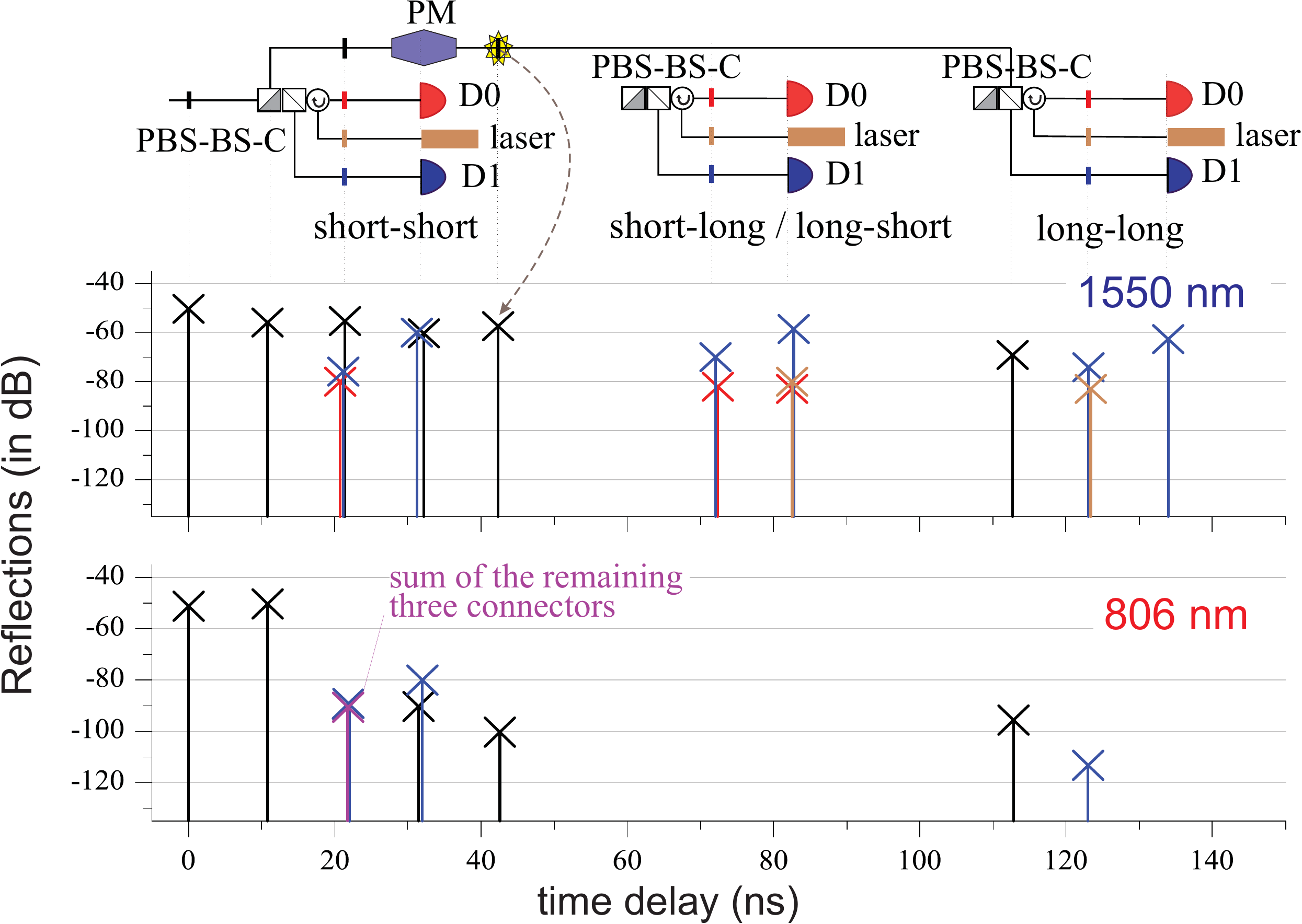}
\caption{Reflection maps of Clavis2-Bob at $1550\,$nm and $806\,$nm, as seen from Bob's entrance. Reflections from several components close in time are color-coded. Reflections not shown were below the OTDR sensitivity (about $-83\,$dB at $1550\,$nm and $-96\,$dB at $806\,$nm). However, some important reflections below the sensitivity limit at $806$ nm were estimated by combining several measurements on parts of Bob. The reflection level of the connectors could depend significantly (maximum variation:$\,3\,$dB) on the cleanliness of the connectors and mating sleeves. In the scheme, small filled rectangular blocks represent FC/PC connectors with curved polished surfaces; PM: phase modulator, D0 and D1: avalanche photodiodes; PBS-BS-C: optical assembly of polarizing beamsplitter, 50/50 beamsplitter, and circulator. OTDR model: Opto-Electronics modular picosecond fiber-optic system.}
\label{fotdr}
\end{figure}
Figure~\ref{fotdr} illustrates two of them; the traces for 1310 nm and 1550 nm were found to be quite similar. Due to the polarizing beamsplitter at Bob's entrance (the PBS in the PBS-BS-C assembly), most of the reflection levels depend greatly on the polarization of the probe light. This polarization was set to maximize the reflection from the closest connector of the PM (see star-like shape). As indicated, the back-reflected pulse would exit Bob around 43 ns after the arrival of the THP into Bob; $t_{B\rightarrow E} - t_{ E\rightarrow B}\sim 43$ ns. The corresponding back-reflection level is around $-57\,$dB. By sending a THP, say with a mean photon number $\mu_{E\rightarrow B} = 2 \times 10^6$, Eve would get a back-reflection $\mu_{B\rightarrow E} \approx 4.0$, i.e.,\ with just four photons on average.
\subsection*{Measurement of the back-reflected pulse}
Per se, any physical property in the back-reflected pulse that provides a clue of Bob's modulation suffices, and governs Eve's measurement technique. If Eve uses a coherent laser operating at wavelength $\lambda_E$ to prepare the THP, the state of light in the back-reflected pulse can be approximated by a weak coherent state $\ket{\alpha}$. The phase $\varphi_E$ = $Arg(\alpha)$ depends on $\lambda_E$, e.g.,\ if $\lambda_E = \lambda_{AB} \sim 1550$ nm, and Eve launches the THP so that both the onward and back-reflected pulse make a pass through the PM while it is active, then $\varphi_E \cong \varphi_B+\pi+\varphi_B = \pi$ or $0$. The objective then simplifies to discriminating between two weak coherent states having the same amplitude $|\alpha|$ but opposite phase, which can succeed with a probability $1 - e^{-|\alpha|^2}$ at most (which is the probability that the state $\ket{\pm \alpha}$ is not projected onto the vacuum state). Assuming the aforementioned case with $|\alpha|^2 \equiv \mu_{B\rightarrow E} \approx 4.0$, the maximal success probability is $98.2\%$. This unknown phase may be probed interferometerically with either a (bright) local oscillator followed by a homodyne detector, or an attenuated coherent state (the same level as $\mu_{B\rightarrow E}$) and a pair of single-photon detectors. 
\subsection*{Avoiding discovery by Bob (or Alice) and other constraints}
Raising $\mu_{E\rightarrow B}$ would yield more photons for the measurement, allowing for a better phase discrimination, but how do these bright pulses affect the other components in the QKD system in general? An oddly-behaving component is a signature that could lead to Eve's discovery, so this issue is quite central to the success of Eve's attack. 

Bob uses a pair of avalanche photodiodes (APDs) operated in gated mode\footnote{Gate width for Clavis2 system is $\approx 2.0$~ns \cite{Lydersen2011}, and gate period is $1/f_B$= 200~ns.} to detect the legitimate photonic qubits from Alice. Eve's bright pulses, even if timed to arrive outside the detection gate, tend to populate carrier traps~\cite{AfterPulPapers,Wiechers2011} in the APD. This ensues in an afterpulsing effect: traps exponentially decay by releasing charge carriers that may stimulate avalanches of current, or \emph{afterpulses}, in the onward gates. These afterpulses increase the dark count rate, i.e., result in higher number of false clicks in the APDs. Due to this, the quantum bit error rate (QBER) incurred by Alice and Bob at the conclusion of the key exchange will naturally be higher. 
Eve's objective is to make sure that the QBER does not cross the `abort threshold' (e.g.,\ around $8\%$ in Clavis2~\cite{Jain2011}) as that would fail her eavesdropping attempt. Moreover, as characterized in the so called after-gate attack~\cite{Wiechers2011}, if the brightness $\mu_{E\rightarrow B}$ exceeds a certain threshold, then for a THP arriving a few ns after the gate, the APD may register a click with high probability for that particular slot. Since Eve wants to \emph{merely} read the state of the phase modulator via a Trojan-horse pulse, she must constrain the brightness of this pulse to avoid an undesired click in Bob's APDs in the attacked slot. This imposes an upper limit on $\mu_{E\rightarrow B}$, which is $\sim 2\times 10^6$ for our system~\cite{Wiechers2011}. 

As the afterpulsing is strongly dependent on the brightness $\mu_{E\rightarrow B}$ and frequency of attack $\textit{f}_{Eatt}$ (which may be lower than $\textit{f}_{B} = 5\,$MHz), Eve would like to attack with the dimmest-possible THPs. The lower limit is mainly decided by the probability of success in discerning Bob's modulation, i.e.,\ how well can Eve's measurement apparatus perform as $\mu_{B\rightarrow E}$ falls in the few-photon regime. Reducing $\textit{f}_{Eatt}$ implies Eve probes only a fraction of the slots that eventually contribute to the raw key formation: she can then possess only a partial amount of knowledge of the raw key. This must therefore be high enough to ensure a positive leakage of information at the end of the protocol, i.e.,\ after Alice and Bob have distilled the secret key by estimating Eve's information and destroying it by means of privacy amplification.
\subsection*{Suitable wavelength for attack}
The behaviour of most optical components is a function of wavelength. The attenuation through fibers and back-reflectance of the connectors may also vary with wavelength. The notable differences between the OTDR traces at $808$ nm and $1550$ nm, shown in figure~\ref{fotdr}, is a testimony to this fact. 

Ideally speaking, to characterize a QKD system, one should perform individual OTDR measurements over a large spectral range that could prove feasible for mounting Trojan-horse attacks. However, identifying such a range is not easy. Moreover, it requires an OTDR system with a tunable source as well as a detector with a high sensitivity over the complete range. This may not be possible in practice. Nevertheless, we made some simple measurements using a photonic crystal fiber based supercontinuum source~\cite{Dudley2006}. The primary focus of these measurements, the details of which will be discussed elsewhere, was to examine the spectral behaviour of Bob's PM in conjunction with its input and output connectors. Fortunately for the QKD system, we did not find any reflection peaks that could have aided Eve. In fact, based on the OTDR and supercontinuum results, the optimum attack wavelength seems to be $\sim 1550$ nm. 
\section{Phase readout experiment}\label{sec:pr}
\subsection*{Eavesdropping setup}
\begin{figure}
\centering
\includegraphics[scale=0.40]{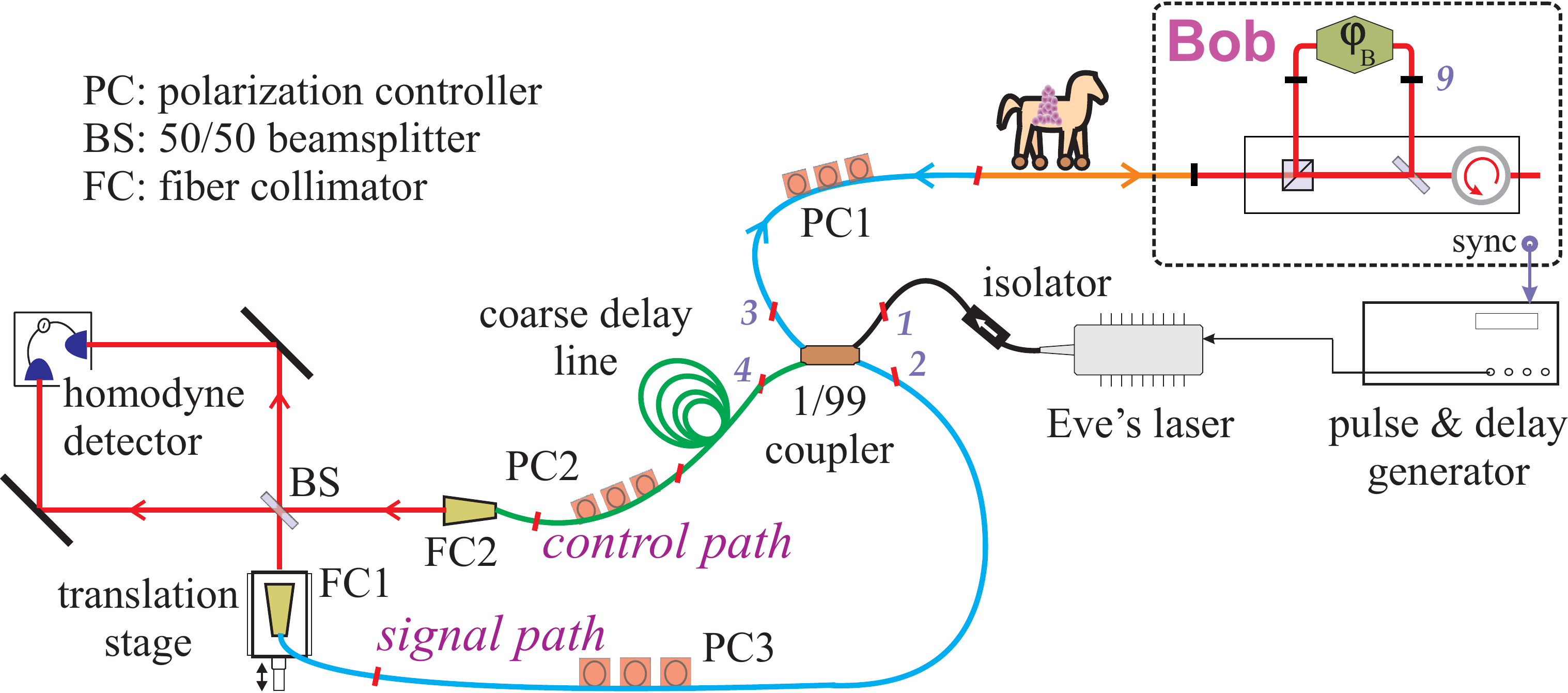} 
\caption{Schematic of a Trojan-horse eavesdropper. Some components in Bob are not shown to avoid cluttering. To synchronize to Bob's modulation cycle, we used an electronic sync signal as shown. In an actual attack, Eve can use the method explained in section~\ref{tsndTHP} (also see the explanation of the attack strategy in the appendix).}
\label{fimplmntn}
\end{figure}

Here we describe our implementation of a proof-of-principle Trojan-horse attack. Figure~\ref{fimplmntn} shows the schematic of the apparatus used for reading out the unknown phase by means of homodyne detection.
For this, we disconnected Bob from Alice. A pulse \& delay generator (Highland Technology P400) was synchronized to Bob and drove Eve's laser at a repetition rate $\textit{f}_{Eatt} = 5$ MHz. An optical isolator was employed to protect Eve's laser from reflections. Using a 50/50 (later replaced by a 1/99) coupler, the Trojan-horse pulses were directed into Bob from port 3. The polarization of these THPs was optimized using PC1 so that the power at the FC/PC connector (port 9, inside Bob) after the PM was maximum. 

A long fiber patchcord of an appropriate length was spliced and added to the other arm of the coupler at port 4. The relative path difference between the back-reflected pulse (\textit{signal} path) and the local oscillator pulse (\textit{control} path), as observed at the 50/50 beamsplitter of the homodyne detector, was adjusted to achieve the maximum interference visibility. The polarization of the \textit{signal} (\textit{control}) pulses at the outcoupler FC1 (FC2) could be controlled by PC2 (PC3). Using P400, the laser delay, i.e.,\ $t_{E\rightarrow B}$ was changed so that the input pulse traveled through Bob's PM while the PM was activated. The optical pulse width, and therefore the mean photon number per pulse, could be fine-tuned by changing $\tau_{Eatt}$, the driving pulse width in P400.
\subsection*{Results}
As mentioned before, Clavis2 operates the quantum key exchange in \emph{frames} that are $215\, \upmu$s long, containing $N_f = 1075$ modulations or slots repeating every $0.2\, \upmu$s. We configured the oscilloscope to capture the output voltage of the homodyne detector and the phase modulator voltage (obtained via an electronic tap placed inside Bob) in a single-shot acquisition mode lasting $250\, \upmu$s. 
\begin{figure}
\centering
\includegraphics[width=0.9\textwidth]{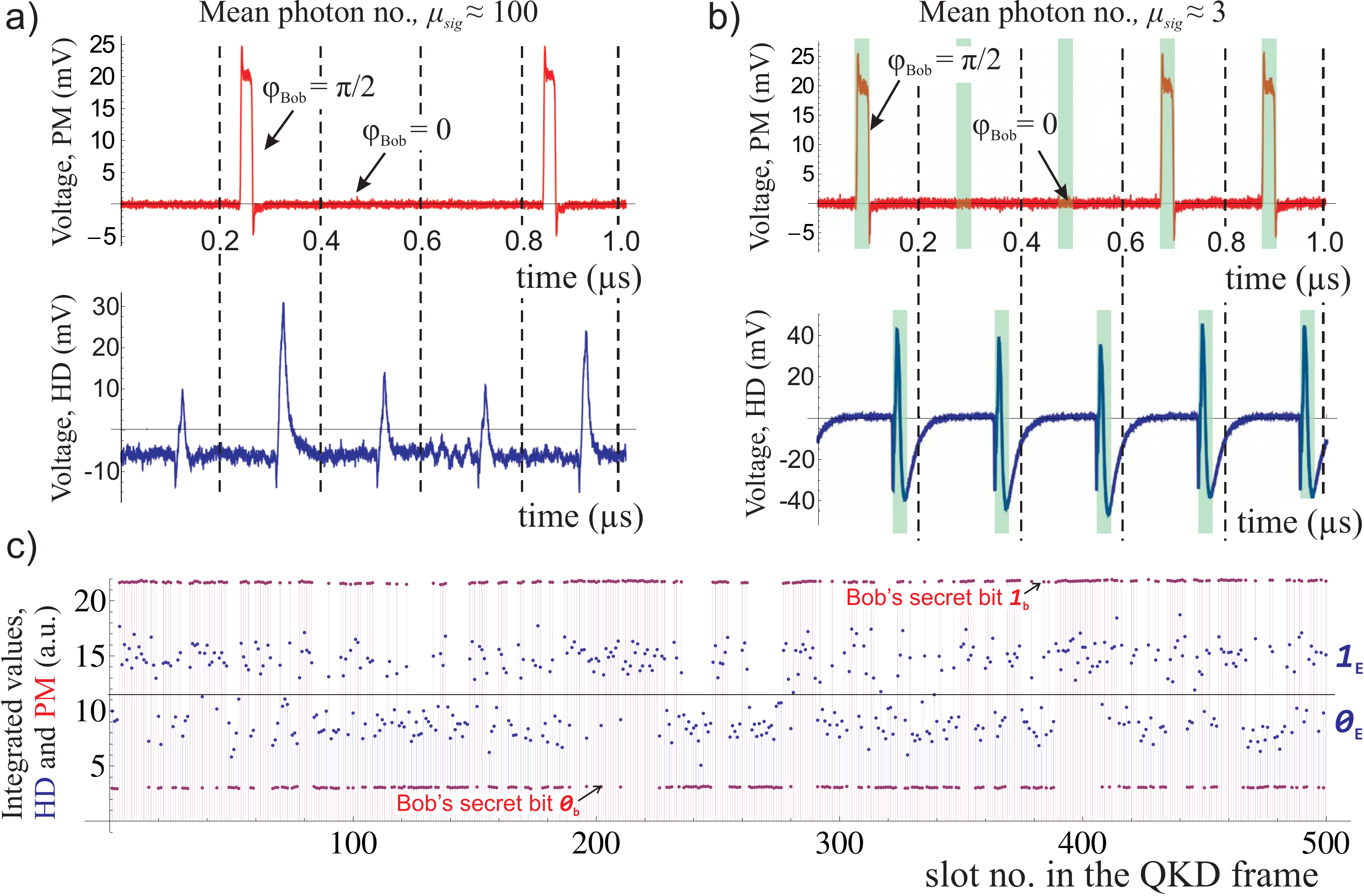} 
\caption{Results of phase readout. (a) Traces of Bob's randomly-chosen phase modulation (in red) and the output of Eve's homodyne detector (in blue) for a sequence of 5 arbitrarily chosen slots. The measurement was performed at $\mu_{sig} \approx 100$ and $\textit{f}_{Eatt} = 5$ MHz. The correlation between Bob's modulation and Eve's homodyne pulses can be easily discerned. (b) Same as in (a) but with $\mu_{E\rightarrow B}$ reduced so that $\mu_{sig} \approx 3$. The major pulse shape observed at the homodyne detector (HD) output arises from the slightly-imperfect subtraction of the LO. The signal is nevertheless easily extracted by integration over a time-window (denoted by green shaded rectangle). Thus, in each $200$ ns slot, a single value each for the random phase modulation and HD pulse is calculated. (c) Series of 500 such integrated values, shifted by an arbitrary constant merely to aid visual discrimination. Using an appropriate threshold (black horizontal line), Eve's estimation of Bob's bit $\texttt{0}_b$ or $\texttt{1}_b$ in a given slot is correct in $>90\%$ cases.}
\label{fphrd}
\end{figure}
Figure~\ref{fphrd} shows the time traces of Bob's randomly-chosen phase modulations and the output of Eve's homodyne detector for 5 arbitrarily chosen slots in two different configurations. The first one (with a 50/50 coupler and $\tau_{Eatt}=3.3$ ns) had mean photon numbers $\mu_{\rm LO} \approx \mu_{E\rightarrow B} = 10^8$ resulting in a mean photon number $\mu_{sig} \approx 100$ of the back-reflected pulses in the signal arm of the homodyne detector. In this case, the discrimination is quite apparent as illustrated in figure~\ref{fphrd}(a); in fact, using peak-to-peak values as a measure, correlations above $99\%$ were easily obtained when measured over entire Clavis2 frames.

We then replaced the 50/50 coupler with a 1/99 coupler and obtained $\mu_{E\rightarrow B} \leq 1.5\times10^6$ at $\tau_{Eatt}=2.6$ ns. In this case, illustrated in figure~\ref{fphrd}(b), the mean photon number $\mu_{sig} \approx 3$ while the LO had slightly higher power than before, $\mu_{\rm LO} > 10^8$. We also confirmed that a slot attacked with the Trojan-horse pulse never experienced a click (except due to a dark count)~\cite{Wiechers2011}. A direct discrimination may not be evident by eye, however, after integrating the homodyne pulses over a suitably chosen time-window every slot, we obtained correlations above $90\%$. This is explained further in the caption, and the corresponding output for 500 slots is depicted in figure~\ref{fphrd}(c). 

Both theoretical and experimentally demonstrated discrimination probability is above $90\%$, and in section~\ref{sec:rnd} we shall discuss a few techniques that can increase it further. To simplify our simulation, we assume from hereon that a Trojan-horse pulse with $\mu_{E\rightarrow B} \sim< 2\times10^6$ can always accurately read the state of Bob's PM in each slot. Finally, note that due to fluctuations, the global phase drifts on frame-to-frame basis, but Eve can always suitably craft her LO to homodyne another back-reflection which passed through Bob's PM outside the modulation width, i.e.,\ when the PM is inactive. This effectively allows to set her reference to $\varphi_B = 0$. Also, such phase drifts are typically in the few-kHz regime which is of the same order as the frame rate in Bob. 
\section{Eve's attack strategy simulation}\label{sec:attstrat} 
To know the entire modulation sequence in Bob, Eve would have to attack the QKD system with $\textit{f}_{Eatt} = 5$ MHz which would result in a tremendous amount of afterpulsing in Bob's APDs even when $\mu_{E\rightarrow B} \sim 2\times10^6$ is chosen. A straightforward attack is clearly not possible. In this section, we devise an attack strategy that may still allow Eve to probe Bob's PM frequently enough to obtain more raw key than Alice and Bob estimate her to possess during the calculation of the secret key fraction~\cite{Branciard2005}. Neither is the expected detection rate of Bob severely affected, nor the QBER crosses the abort threshold. In other words, \emph{a non-zero portion of the final secret key is leaked to Eve without her being discovered}. 

To motivate the basic idea of the strategy, note that it makes sense to probe the modulation in a slot if Bob, with a high probability, eventually obtains a valid detection in that slot. Conversely, if a slot has a very low probability of being registered by Bob, probing that slot is not only a waste but also the afterpulsing -- due to Eve's bright pulses -- unnecessarily increases the QBER. By manipulating the photonic frame, i.e.,\ the train of $N_f = 1075$ legitimate weak coherent pulses (WCPs) returning from Alice to Bob, Eve can control the timings of detection events in Bob. For this purpose, she may either (i) use a low-loss channel to transfer the photon(s) in a WCP from Alice to Bob and increase the chance of a click in that given slot, or (ii) block the WCP entirely to decrease it. She multiplexes Trojan-horse pulses on (a subset of) the former slots as depicted in figure~\ref{fpnp}(a) while keeping her laser shut in the latter slots.

Since the mean photon number of the WCPs arriving in Bob is rather low, a major chunk of the slots would actually contain $0$ photons, and obviously cannot result in a detection event in Bob. Eve may increase her chance of attacking a slot, that eventually yields a valid detection event, by sending a set of consecutive Trojan-horse pulses, here called an \emph{attack burst} with length $N_{ab}$. However, this burst would also cause a large amount of afterpulsing -- noticeable even a few slots after its application. Eve's remedy to this is based on the fact that a successful click causes a \emph{deadtime} in Bob's APDs. During the attack burst, Eve therefore tries to impose a deadtime in Bob from Alice's photons to \emph{mask} the afterpulsing. To achieve that, she uses the low-loss channel to transfer the $N_{ab}$ slots to Bob to increase the photon detection probability. 

Since $N_{ab}$ can obviously not be too large, the deadtime imposition (which results in a withdrawal of $N_{dt} = 50$ gates in Clavis2) may not always work during the attack burst. Therefore, Eve also transfers another set of $N_{ss}$ slots on the low-loss channel, called the \emph{substitution sequence}, to keep the photon detection probability high after the attack burst as well. We emphasize that Eve does not add any Trojan-horse pulses during the substitution sequence. 

In this scenario, the detection clicks in Bob's APDs due to Alice's photons (sent over the low-loss channel in $N_{ab}+N_{ss}$ slots) compete with those from the afterpulses: the former may mask the latter, effectively lowering the error probability. Finally, another optimization for Eve would involve drastically decreasing the detection probability before these $N_{ab}+N_{ss}$ slots -- otherwise, a click in a slot before the attack burst slots would result in the burst being encompassed in a deadtime, yielding no benefit to Eve. By extinguishing a certain number of the WCPs (denoted as \emph{extinguished length} $N_{el}$), she may reduce these chances. 
\begin{figure}
\centering
\includegraphics[scale=0.36]{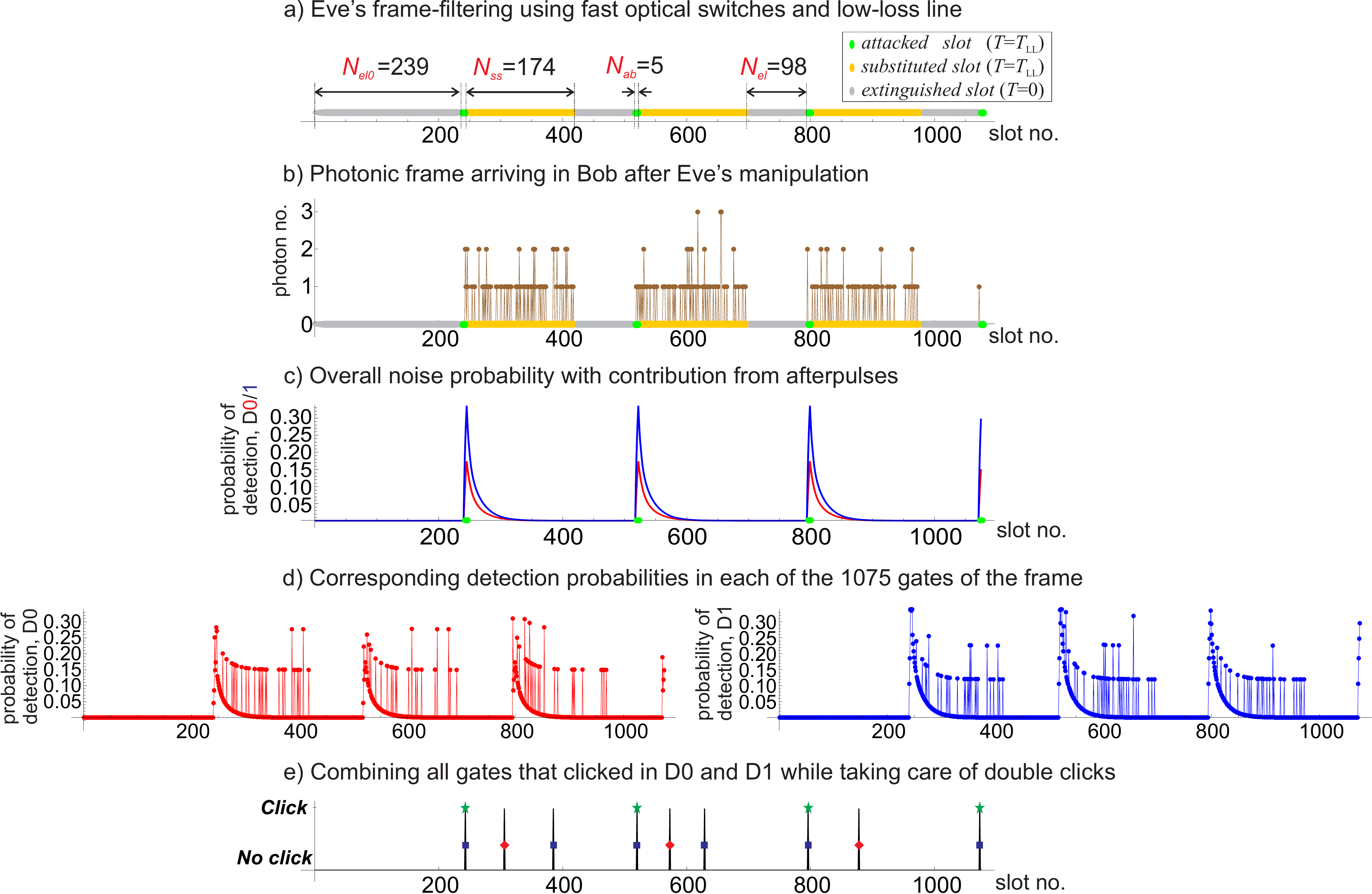}
\caption{Simulating the effect of Eve's attack on the QKD protocol operation. (a) Eve manipulates a frame sent by Alice to Bob using the strategy described in the main text (more details in the appendix). (b) Bob receives a `filtered' frame, as the \emph{effective} channel transmission is $T=T_{\rm LL}$ for all $N_{ab}$ (attack burst) and $N_{ss}$ (substitution sequence) slots, and $T=0$ for $N_{el}$ (extinguished length) slots. We assumed $T_{\rm LL}=0.9$ in the present case. (c) Characteristic exponential decay of probabilities due to afterpulsing in both D0 and D1 (red and blue) may be visualized after the attack-bursts. (d) Final detection-probability patterns for D0 and D1. (e) Subsequent click pattern just like in figure~\ref{fphnNdet}(e); out of the 9 slots (3 in D0 and 6 in D1, indicated by rotated-red and straight-blue squares, respectively) where clicks occurred, Eve knows the basis choice of Bob in 4 of them (indicated by green star).}
\label{fphnNdetEve}
\end{figure}
Thus, her attack pattern can be thought of as a repetition of the triad $\{N_{el}$, $N_{ab}$, $N_{ss}\}$, as illustrated by an example in figure~\ref{fphnNdetEve}(a). 
\subsection*{Evaluating the QKD frame manipulation}
In the appendix, we describe a specific construction of Eve's strategy using fast optical switches~\cite{fos} and low-loss channels for manipulating the QKD frame as explained above. Due to this manipulation, Bob receives photons from Alice only during the attack bursts and substitution sequences. This is apparent in figure~\ref{fphnNdetEve}(b); see the thick yellow and green segments. Also, due to the afterpulses emanating from the attack burst slots, the dark noise is not uniform throughout the frame. The overall noise probability in the $l^{th}$ slot is given by $n_{j}(l) = d_{j} + a_{j}(l) - d_{j}\times a_{j}(l)$ for $j = 0$ and $1$, and is shown in figure~\ref{fphnNdetEve}(c). In this expression, $d_{0/1}$ represents the dark noise probability per gate for D0/D1. The function $a_{j}(l)$ is computed by summing together the contributions of all previous afterpulses until the $l^{th}$ slot; this is explained in more detail in Ref.~\cite{Wiechers2011}. Table~\ref{tDetparams} lists all the parameters for calculating the function $n_{j}(l)$.

After considering both the photonic input and noise figure, we can evaluate the final detection probabilities $p_{j}(l) = s_{j}(l) + n_{j}(l) - s_{j}(l)\times n_{j}(l)$ for the entire frame, as shown in figure~\ref{fphnNdetEve}(d). We explain the derivation of $s_{j}(l)$ and modelling of the click events in D0 and D1 based on Bernoulli trials in the appendix. Figure~\ref{fphnNdetEve}(e) illustrates the clicked gates found after taking double clicks and deadtime imposition into account. Note that while Eve attacked only 20 out of 1075 slots, she knows Bob's basis choice in 4 out of 9 slots that are going to be used in the formation of the raw key.\label{detmscAttck} 

The QBER incurred by Alice and Bob is strongly dependent on the combination $\{N_{ab}$, $N_{ss}$, $N_{el}\}$ used by Eve during the operation of the QKD protocol. The quantum channel transmission $T$ and low-loss line transmission $T_{\rm LL}$ directly influence the photon number statistics $\mu_{SARG04}$ in Alice and the observed detection rate $\gamma_B$ in Bob, and also indirectly affect both the QBER and Eve's actual correlations $I^{\rm act}_E$ with the key shared by Alice and Bob after error correction. For instance, long and frequent attacks (larger $N_{ab}$ and smaller $N_{ss}$, in a relative sense) yield high $I^{\rm act}_E$ but also high QBER. Similarly, a large $N_{el}$ preceding an attack burst may effectively increase $I^{\rm act}_E$ as the attacked slots have lesser chances of being inside a deadtime period, but this may also decrease $\gamma_B$. And a high $T_{\rm LL}$ naturally implies higher $\gamma_B$, and perhaps lower QBER because the dark noise is effectively decreased, however $T_{\rm LL}$ cannot exceed~1. 
\subsection*{Classical processing and optimizing the simulation}\label{sec:ocp}
Let us first briefly recapitulate some essential information from the previous pages. In section~\ref{sec:pr}, we experimentally demonstrated the readout of Bob's phase modulator with a high accuracy. However, we also found that frequent Trojan-horse pulses would result in a huge afterpulsing in Bob's APDs which would reveal Eve's presence easily. In this section, we devised an intuitive strategy in which Eve manipulates the frame-based communication of Clavis2 and attacks (with Trojan-horse pulses) only a small but carefully-chosen subset of the slots in a frame. If Eve simultaneously ensures that
\begin{enumerate}
	\item the QBER $q$ does not cross the abort threshold ($q < q_{\rm abort}$), 
	\item the portion of the raw key Eve actually knows is more than whatever Alice and Bob estimate based on the security proof ($I^{\rm act}_E > I^{\rm est}_E$), and 
	\item the deviation of the observed detection rate $\gamma^{\rm obs}_B$ from the expected value in Bob $\gamma^{\rm exp}_B$, given by $\delta_B = \left|1-\frac{\gamma^{\rm obs}_B}{\gamma^{\rm exp}_B}\right|$, is within tolerable limit ($\delta_B \leq \delta^{\rm max}_B$), 
\end{enumerate}
then her strategy succeeds. For satisfying these requirements, one needs to find an optimal attack combination. We simulated different combinations $\{r$, $N_{ab}$, $N_{ss}$, $N_{el}\}$; with the new variable $r \leq 1$ denoting the fraction of frames subjected to the Trojan-horse attack. To elaborate, if $r=0.8$, Eve randomly chose $80$ out of $100$ frames to attack with the pattern imposed by a specific triad $\{N_{ab}$, $N_{el}$, $N_{ss}\}$ in the manner shown in figure~\ref{fphnNdetEve}, while the remaining $20$ passed to Bob normally (in the manner shown in figure~\ref{fphnNdet}). 

Due to probabilistic elements in the simulation, each run was performed for $n_{\rm sim}=10000$ frames to minimize stochastic fluctuations. In each run, slots that yielded clicks were collated and the average number of clicks per frame $\gamma^{\rm obs}_B =$ (total clicks)/$n_{\rm sim}$ was calculated. A basis reconciliation procedure, as per the specifications of SARG04~\cite{Scarani2004a,Branciard2005}, was then performed on the collated slots. This provided us with the incurred QBER $q$ and the fraction of \emph{valid} slots\footnote{I.e.,\ the slots kept by both Alice and Bob after the basis reconciliation.} in which Eve knows the secret bit. From the former, we can calculate the leak due to error correction (EC) $leak_{\rm EC}$ then use it with the latter to bound Eve's correlations $I^{\rm act}_E$ with the error-corrected key. In particular, we assumed EC to work in the Shannon limit, i.e.,\ $leak_{\rm EC} = h(q)$, with $h(x) = -x\log_2(x)-(1-x)\log_2(1-x)$ being the binary entropy. 

To calculate the amount of privacy amplification that Alice and Bob do in SARG04 protocol, we evaluated the expression $I(A:E)$ derived in Ref.~\cite{Branciard2005} (equation (88) therein); this provides $I^{\rm est}_E$ essentially. The derivation considers eavesdropping strategies applicable against SARG04 when Alice employs an attenuated laser instead of a single-photon source. The final expression is obtained while optimizing and lower-bounding the secret key fraction attained by Alice and Bob. 

One element considered in the calculation of $I^{\rm act}_E = I(A:E)$ is \emph{preprocessing}: a classical operation performed by Alice at the commencement of QKD that reduces both Bob's and Eve's information, but in a more inimical manner for the latter than the former~\cite{Kraus2005a,Branciard2005}. Although Ref.~\cite{Branciard2005} concludes that preprocessing in SARG04 helps Alice and Bob only in a very specific regime, it does not explicitly state that preprocessing should be avoided in other regimes. Since security proofs generally consider attacks that maximize $I(A:E)$ instead of $I(B\!:\!E)$, the use of preprocessing by Alice may expose a vulnerability exploitable via Trojan-horse attacks on Bob. Although preprocessing is not implemented in Clavis2, we consider a case here to highlight the vulnerability.

Indicating the degree of preprocessing performed by Alice with a variable $y$, and using all the relevant source, channel, and detector parameters introduced thus far, we calculate $I^{\rm est}_E = 0.4844$ for $y=0$. This implies that Alice and Bob compress almost half of their error-corrected key during privacy amplification. If however, Alice were to use the maximum preprocessing ($y=0.5$), then $I^{\rm est}_E = 0.1106$. Note that the value of $I^{\rm est}_E$ is independent of the incurred QBER. This is due to the fact that the attacks found optimal in the security proof~\cite{Branciard2005} are `zero-error' attacks~\cite{Scarani2009a}. However, $I^{\rm est}_E$ depends on the channel transmission, as also shown in Ref.~\cite{Branciard2005}. The values here are calculated at a fixed transmission ($T=0.25$).
\section{Results and discussion}\label{sec:rnd}
\begin{figure}
\centering
\includegraphics[scale=0.43]{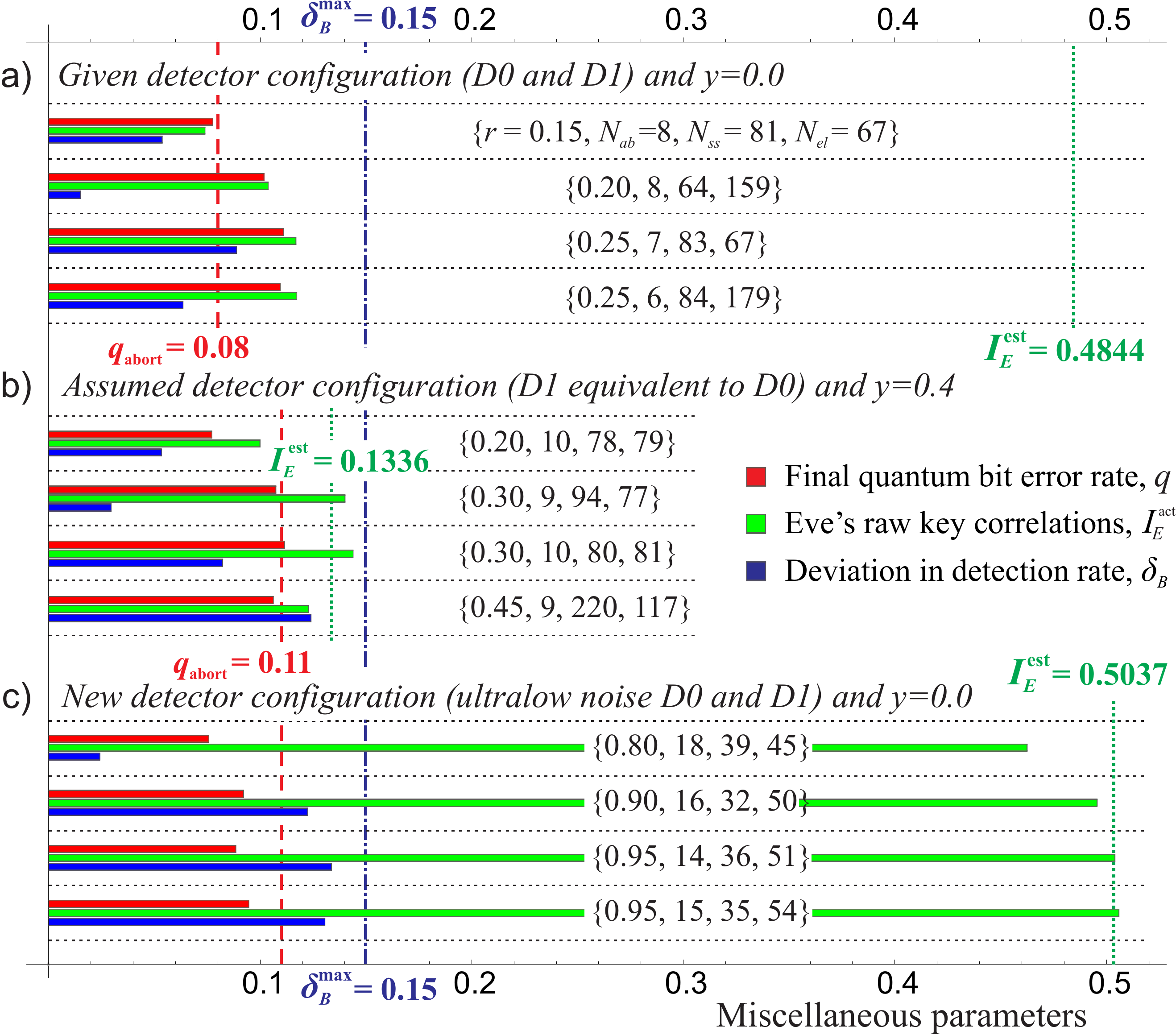}
\caption{Performance of the simulated attack strategy in three different scenarios. The QKD system aborts the protocol when the QBER $q$ crosses a threshold $q_{\rm abort}$ (dashed red line) or the absolute deviation in the detection rate $\delta_B$ surpasses a boundary $\delta^{\rm max}_B$ (dash-dotted blue line). To break the security under these constraints, Eve's actual correlations with the raw key $I^{\rm act}_E$ must exceed the estimate made by Alice and Bob $I^{\rm est}_E$ (dotted green line). (a) Assuming D0 and D1 with characteristics as that of the Clavis2 detectors (see Table~\ref{tDetparams}) and that Alice does not apply any preprocessing ($y=0$), it seems difficult to satisfy the three conditions: $q < q_{\rm abort}$, $\delta_B \leq \delta^{\rm max}_B$, and $I^{\rm act}_E > I^{\rm est}_E$ simultaneously. (b) Assuming both detectors behaving like D0, some preprocessing ($y=0.4$), and $q_{\rm abort} \approx 0.11$, Eve can breach the security. (c) A QKD system implemented with APDs having high efficiency and low noise is vulnerable to Trojan-horse attack even without the preprocessing loophole. The optimal attack combinations $\{r$, $N_{ab}$, $N_{ss}$, $N_{el}\}$ that produced these results are also listed (see text for details). All parameters and results were computed at $T=0.25$ and $T_{\rm LL}=0.9$.}
\label{fresSim}
\end{figure}

In trying to search for optimal combinations $\{r$, $N_{ab}$, $N_{el}$, $N_{ss}\}$ that satisfy all the requirements listed in the previous section, we could find numerous cases where two of the three conditions were easily satisfied ($q_{\rm abort} \approx 0.08$ and $\delta^{\rm max}_B = 0.15$ for Clavis2), as shown in figure~\ref{fresSim}(a). However, it is clear that below the QBER abort threshold, the final raw key correlations of Eve never surpass the estimate of Alice and Bob, i.e.,\ $I^{\rm act}_E < I^{\rm est}_E$. One reason for the failure is that the detectors, especially D1, in Clavis2 are quite noisy: even without an attack, i.e.,\ with $r=0$, the QBER $q=2.52\%$. Crafting an attack with high $r$ and optimal $\{N_{ab}, N_{ss}, N_{el}\}$ may give Eve sufficiently high $I^{\rm act}_E$ but the incurred QBER $q >> q_{\rm abort}$. 

If we assume Bob's detectors to have the same characteristics as that of D0 (in Clavis2), and that Alice has preprocessing accidentally enabled, then Eve could breach the security for $q_{\rm abort} \approx 0.11$ as shown in figure~\ref{fresSim}(b). 
This is possible because the mutual information between Eve and Bob scales by the same factor (given by $1-y$) as that between Alice and Bob: in particular, at $y=0.4$, Eve can surpass $I^{\rm est}_E = 0.1336$. 

In order to gauge the full power of this attack strategy and the dangers posed by Trojan-horse attacks in general, we optimized the simulation for a Clavis2-like QKD system assumed to be fitted with a pair of APDs having \emph{high efficiency} and \emph{low noise}. To be more precise, we assumed a pair of gated APDs with detection efficiencies $\eta_0 = \eta_1 = 0.25$, thermal dark count probabilities $d_0 = d_1 = 10^{-5}$ per gate, and a cumulative probability of obtaining random click after deadtime period due to afterpulses to be $<10\%$ (refer Table~\ref{tDetparams} for comparison). Note that detectors with similar or even better characteristics have already been reported~\cite{Patel2012,Walenta2012,Restelli2012,Korzh2013}, thanks to the recent advances in single-photon detection technology. Alternatively, mechanisms to photoionize the trapped charges through sub-band energy illumination in order to reduce afterpulsing have also been investigated~\cite{Krainak2005}. Therefore, it is quite reasonable to expect such characteristics in the next-generation gated APDs in Clavis2 or recently-manufactured QKD devices. In such QKD systems, not only can Eve attack more often, but also expect detections from photons to exceed those from afterpulses.

Figure~\ref{fresSim}(c) shows some optimized attacks ($I^{\rm est}_E = 0.5037$ for the new detector parameters and no preprocessing) that satisfy all the three conditions. In particular, the positive leakage $I^{\rm act}_E - I^{\rm est}_E$, which is likely to be higher when preprocessing is also used, implies that the security of the QKD system would be breached.

At lower channel transmission values ($T<0.25$), attack regimes with a positive leakage of final secret key may be found by means of more exhaustive optimization of the simulation. At higher transmission values ($T>0.25$), Eve's attack should have better chances of succeeding because Alice's quantum states have more photons on average, which raises the photonic detection probability (effectively suppressing the afterpulsing probability) in Bob. However, the calculation of $I^{\rm est}_E$ in the security proof~\cite{Branciard2005} is valid only for channel lengths above $24$ km, translating roughly into $T<0.33$. More photons from Alice also raise the chances of better photon-number-splitting attacks~\cite{Brassard2000,Jiang2012} which would have to be countered by increasing $I^{\rm est}_E$ in privacy amplification, thereby requiring Eve to work harder.

Nonetheless, it is clear that our attack on a QKD system equipped with less noisy APDs would succeed at least for a range of channel transmissions. Furthermore, a finite amount of preprocessing -- supposed to provide more security to Alice and Bob -- would actually relax the constraints on Eve. Finally, the Trojan-horse strategy could be combined with other hacking strategies, such as the after-gate attack~\cite{Wiechers2011}, to enhance Eve's performance.  
\subsection*{Possible improvements and extensions}\label{sec:imprvmnts}
An optimization over the complete space of all parameters that define the attack strategy is out of the scope of this work, but a powerful adversary can easily do so and is likely to find a new set of parameters with better attack performance. 
A possible extension of the strategy is to manipulate the frames from Bob to Alice as well: more precisely, to replace the legitimate bright pulses in  the slots chosen for the \emph{attack burst} with even brighter ones. This would increase the chances that these slots eventually yield valid detections in Bob. Unfortunately, an increased optical power, even if only for a few pulses in the frame, portends a risk for Eve because the monitoring detectors in Alice may raise an alarm. However, if the monitoring system in Alice either does not function properly, or can be fooled \cite{Sajeed2014}, then this method holds a lot of promise. 

Yet another attack optimization is non-demolition measurement~\cite{Braginsky1996,Xiao2008} of the photon numbers of the WCPs exiting Alice. Using it, Eve can simply withhold her attack in the slots that contain $0$ photons. This would reduce the dark counts (from afterpulsing), yet effectively increase her correlations with the raw key.  
Finally, with regards to the attack setup shown in figure~\ref{fimplmntn}, Eve could:
\begin{itemize}
\item gather more information (per phase modulation) by suitably tweaking her LO to homodyne \emph{multiple} back-reflections and improve the quality of the phase readout, 
\item periodically track the phase drift in her setup and adjust the relative phase between the signal and LO, e.g.,\ by using an extra phase modulator in the LO arm, to always read out at an optimal phase difference, and/or
\item enhance the success rate of discrimination by using better quantum measurement strategies~\cite{Wittmann2008} and post-processing techniques, e.g.,\ taking the difference of consecutive pulses and then integrating over the properly-chosen time window. 
\end{itemize}
These methods would facilitate $\sim 100\%$ correlations between Eve's homodyne output and Bob's modulation (see figure~\ref{fphrd}) while relaxing the brightness requirement, i.e.,\ $\mu_{E\rightarrow B}$ may be lowered, thus bringing down the afterpulsing probability. Another way to achieve the same goal would be to employ longer wavelengths to attack (as the afterpulsing response of the APDs is conjectured to be lower) and/or to depopulate the traps by means of photoionization. Eve could try to use $\sim 1700\,$nm for her Trojan-horse pulses to reduce afterpulsing. A CW illumination at a longer wavelength $\sim 1950\,$nm may depopulate the traps (created due to the Trojan-horse pulses at some other wavelength) by means of photoionization~\cite{Krainak2005}.

The attack setup shown in figure~\ref{fimplmntn} can be used virtually against any kind of QKD system, including CVQKD devices~\cite{Jouguet2013,Khan2013a}; it only needs a careful delay and polarization control and interferometric stability. By integrating a variable optical delay line and splicing the different components, it could readily be assembled into a portable setup. Finally, the strategy detailed above can also be attuned to attack entanglement-based QKD systems that may not have proper safeguards against Trojan-horse attacks. More significantly, it may be used even to break the BB84 protocol in such cases. 
\subsection*{Countermeasures}
Experimentally speaking, isolators and wavelength filters have been the most suitable countermeasures against Trojan-horse type attacks for one-way QKD systems \cite{Vakhitov2001}. While the former cannot be used in a two-way QKD system like Clavis2, the latter can certainly be useful. In a related context, one must also scrutinize (high and unwarranted) back-reflections from the interfaces inside the QKD device that could pose risks as explained in section~\ref{sec:thry}. With such analysis, it might be possible to incorporate Trojan-horse attacks into theoretical security proofs and neutralize them by correct levels of privacy amplification. Moreover, security proofs should also carefully examine and quell the undesired effects of preprocessing. Some technical countermeasures specifically for the Clavis2 system could be: 
\begin{itemize}
	\item installing a watchdog detector with a switch at the entrance of Bob that randomly routes a small fraction of incoming signals to this detector,
	\item opening the door for Eve for a smaller time duration, i.e.,\ reducing the width of phase modulation voltage pulse, and 
	\item monitoring Bob's APDs in real time~\cite{Silva2012}.
\end{itemize}
Except the watchdog detector countermeasure, all others require modifications only in the electronic control system and hence are recommended.

Note that Bob's vulnerability to the Trojan-horse attack only arises because the SARG04 protocol is used. For BB84 (including its decoy-state version), interrogating Bob's modulator gives Eve no advantage~\cite{Vakhitov2001}, except when this is used to counterattack the four-state patch to the detector efficiency mismatch attacks~\cite{Makarov2006,Qi2007}. However both BB84 and SARG04 are vulnerable to interrogating Alice's modulator.
\section{Conclusion}
In conclusion, we have demonstrated the operation of a setup to launch a Trojan-horse attack on a commercial QKD system from ID~Quantique. Our objective is to read the state of the phase modulator in Bob to break the SARG04 protocol. We have shown that this phase readout can be performed in real-time with a high success rate, and analyzed various constraints and problems in mounting a full attack on the system. These arise mainly due to the afterpulsing noise induced in the single-photon detectors of Bob by the bright Trojan-horse pulses from Eve. We have devised and numerically modeled an attack strategy to keep the overall QBER (which increases due to the afterpulsing noise) below the abort threshold, while allowing Eve to obtain the maximum possible correlations with the raw key. Although, on our Clavis2 system, this does not exceed the theoretical security estimate that Alice and Bob make about Eve's correlations, we have shown that similar or future QKD systems with less-noisy detectors may be hacked using this strategy. We have also proposed some mechanisms to improve the performance of the attack. With some simple modifications, our attack setup and strategy could be applied against many other quantum cryptographic implementations, including entanglement-based, continuous-variable, and measurement-device-independent QKD systems. Finally, we have proposed both general and specific countermeasures that can be easily adopted in most QKD systems. 
\ack{We would like to thank Matthieu Legr\'e from ID~Quantique, Denis Sych, Christoffer Wittmann, and Lars Lydersen for useful discussions. We also gratefully acknowledge Lothar Meier and Adam K\"appel for their assistance in design of electronics. This work was supported by the Research Council of Norway (grant no.\ 180439/V30), Industry Canada, DAADppp mobility program financed by NFR (project no.\ 199854) and DAAD (project no.\ 50727598). E.A.\ acknowledges support from CryptoWorks21. V.M.\ acknowledges support from University Graduate Center in Kjeller.
\bigskip

\appendix
\section*{Appendix}
\setcounter{section}{1}
\subsection*{Operation of plug-and-play QKD}
Here we simulate the operation of the QKD system. A Clavis2 frame consists of $N_f = 1075$ slots spaced $0.2\, \upmu$s apart. This implies $N_f$ optical signals are sent by Bob to Alice in the forward path of the plug-and-play scheme, $N_f$ detection gates are opened by Bob to measure the $N_f$ weak coherent pulses (WCPs) coming back from Alice\footnote{In practice, Bob has an asymmetric interferometer as shown in figure~\ref{fpnp}(b) so an optical signal actually consists of two (unequally bright) pulses. As it does not affect our analysis, we will use `signal' and `pulse' interchangeably to keep the explanation simple.}. Alice attenuates these optical signals properly so that the mean photon number of the WCPs (in the quantum channel) is as dictated by the protocol; for SARG04 the optimal value is $\mu_{\rm SARG04} = 2\sqrt{T}$, where $T$ is the channel transmission~\cite{Branciard2005}. 

By means of a Monte Carlo simulation based on experimental parameters, we modelled the frame-based QKD operation from hereon. We created an array of random positive integers that are Poisson-distributed to mimic (the photon numbers of) a Clavis2 frame exiting Alice. Each pulse in the frame was stochastically subjected to all the relevant transmission or detection events; to be precise, they were modelled by a sequence of Bernoulli trials. For example, if the transmission of the quantum channel is denoted by $T$, then each of the $n$ photons in a pulse at Alice's exit undergoes a Bernoulli trial yielding success/$1$ [failure/$0$] with a probability of $T$ $\left[1 - T\right]$. The total number of photons in a pulse reaching Bob can then be evaluated as the sum of the outcomes of all $n$ trials. Similarly, for a pulse containing $m$ photons impinging on an APD with single-photon detection efficiency $\eta$, a detection click (success) is obtained if at least one of the $m$ Bernoulli trials yielded a $1$. 

Figure~\ref{fphnNdet} charts the different events in Bob: right from the arrival of a photonic frame to the registration of clicks, taking the withdrawal of $N_{dt} = 50$ gates (due to deadtime) into account. 
\begin{figure}
\centering
\includegraphics[scale=0.35]{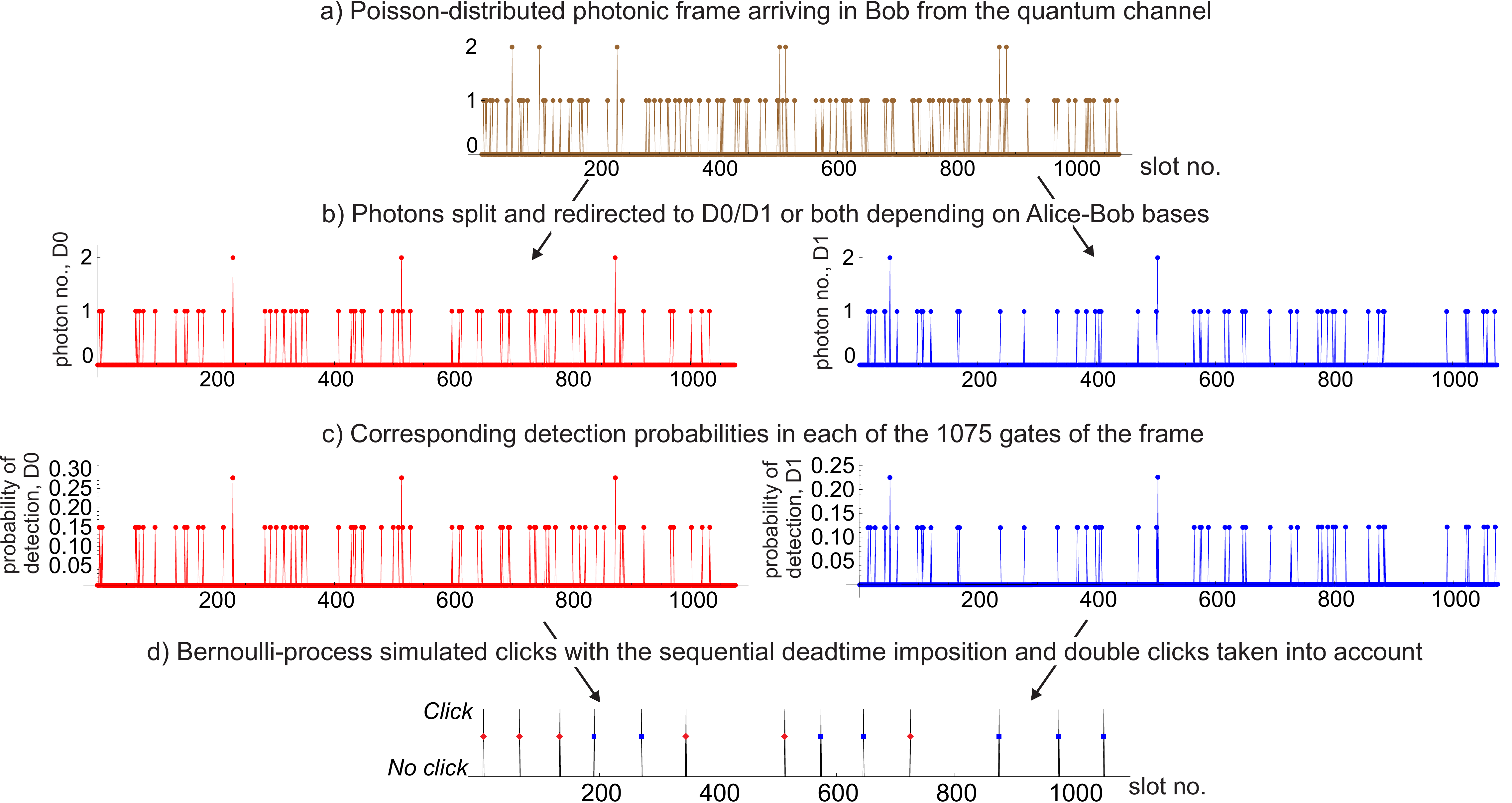}
\caption{Simulation of the physical-layer operation of SARG04 in Clavis2 at channel transmission $T=0.25$. (a) Photon number statistics of the WCP train (mean photon number $\mu_{\rm SARG04} = 1$) that reaches Bob after traversing the quantum channel. In each of the $1075$ slots, Alice randomly prepared one of four states $Z0$, $Z1$, $X0$, $X1$. (b) Bob randomly chose $Z$ or $X$ basis in each slot too; if his basis coincides with the preparation-basis of Alice, all photons in that slot are directed to one of D0 or D1 (depending on Alice's secret bit). For dissimilar basis choice, photons are randomly split across D0 and D1. (c) Resultant detection probabilities for D0 and D1 in each slot/gate; calculation details are given in the main text. (d) Subsequent detection-click pattern (vertical black bars with rotated-red or straight-blue squares).} 
\label{fphnNdet}
\end{figure}
The transmission of the quantum channel connecting Alice and Bob is assumed to be $T=0.25$ (with channel attenuation $\alpha = 0.2$ dB/km, this would imply $\sim 30$ km long channel). The transmission inside Bob is $T_B=0.45$. The total detection probabilities in figure~\ref{fphnNdet}(c) are calculated using $p_{j}(l) = s_{j}(l) + d_{j} - s_{j}(l)\times d_{j}$ for each slot $l \in \left[1,N_f\right]$ and for $j = 0$ and $1$. In this expression, $d_{0/1}$ represents the dark count probability per gate for D0/D1. \label{spdp}
The photonic detection probability is $s_{j}(l) = 1-\left(1-\eta_{j}\right)^{m(l)}$ for $j = 0$ and $1$; here $m(l)$ is the number of photons impinging on a specific detector in the $l^{th}$ slot (shown in figure~\ref{fphnNdet}(b)), and $\eta_{0}$ and $\eta_{1}$ are the single-photon detection efficiencies of D0 and D1, respectively.  
\begin{table}\centering
\newcommand\Ta{\rule{0pt}{2.9ex}}
\newcommand\Tb{\rule{0pt}{2.4ex}}
\caption{Various detection-related parameters in Clavis2. The numerical parameters for the exponential decay due to afterpulses were estimated in Ref.~\cite{Wiechers2011}. The cumulative probability to get a random click after $N_{dt}=50$ gates from afterpulses alone surpasses $80\%$. The subscript $j=0/1$ in a variable affiliates it to D0/D1.\label{tDetparams}}
\begin{tabular}{|c|c|c|}
\hline
\Tb & D0 & D1  \\
\hline
Single-photon detection efficiency, $\eta_j$ \Tb & $0.12$ & $0.10$  \\
\hline
Dark noise probability, $d_j$ \Tb & $1.16\times10^{-4}$ & $3.63\times10^{-4}$  \\
\hline
Afterpulse probability amplitude, $A_{1j}$ \Tb & $3.572\times10^{-2}$ & $10.68\times10^{-2}$\\
\hline
Afterpulse decay constant, $\tau_{1j}$ ($\mu s$) \Tb & $1.159$ & $0.705$  \\
\hline
Afterpulse probability amplitude, $A_{2j}$ \Tb & $2.283\times10^{-2}$ & $5.054\times10^{-2}$\\
\hline
Afterpulsing decay constant, $\tau_{2j}$ ($\mu s$) \Tb & $4.277$ & $3.866$  \\
\hline
\end{tabular} 
\end{table}
Table~\ref{tDetparams} lists the various parameters relevant to the detectors. 
\subsection*{Eve's strategy}\label{syncEve2}
Figure~\ref{fstrtgy}(a) shows a possible full implementation of the Trojan-horse attack described in section \ref{sec:attstrat}, by using off-the-shelf optical switches \cite{fos} and a low-loss line. 
\begin{figure}
\centering
\includegraphics[scale=0.41]{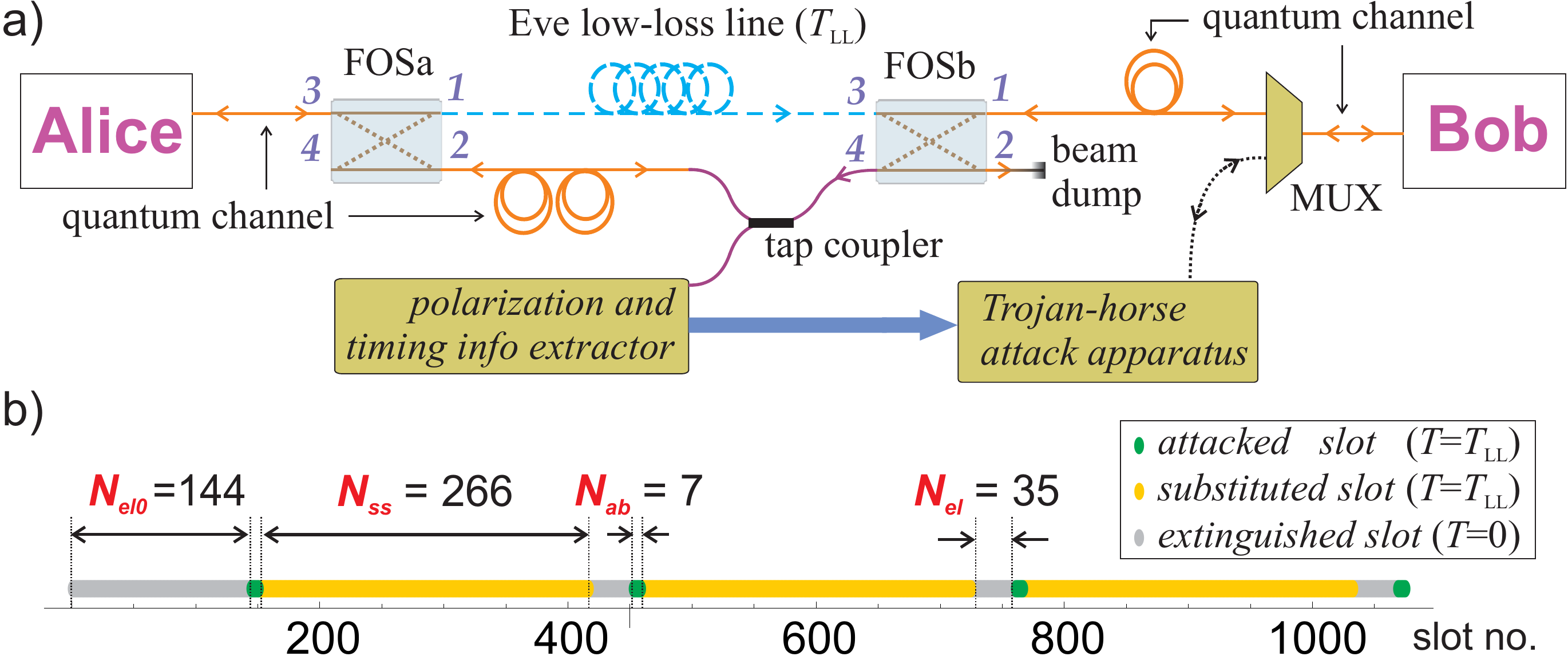}
\caption{Technical implementation details of the frame manipulation strategy. (a) Eve plants two bi-directional $2\!\times\!2$ fast optical switches FOSa and FOSb near Alice and Bob, respectively. The solid orange line represents the quantum channel (normal transmission $T$) containing an optical tap along with the two switches. The dashed cyan line is Eve's highly-transmissive channel which may be implemented by a low-loss delay line. The operational details of the switches during the quantum key exchange are described in the text. (b) In a frame sent by Alice to Bob, Eve diverts all the slots marked in green and yellow (four sets of $N_{ab}$ and three sets of $N_{ss}$, respectively) onto a highly-transmissive channel. The slots marked in grey (three sets of $N_{el}$ and one $N_{el0}$) are blocked. FOS: fast optical switch, MUX: multiplexer, \emph{ab}: attack burst, \emph{ss}: substitution sequence, \emph{el}: extinguished length.}
\label{fstrtgy}
\end{figure}
The switches are connected by two lines: the quantum channel containing an optical tap additionally, and a highly-transmissive channel (with transmission $T_{\rm LL}$). If a slot $l \in \left[1,N_f\right]$ diverted by Eve on the highly-transmissive channel had $n$ photons at Alice's exit, then it has a high chance of having $n$ photons at Bob's entrance too. The low-loss line with the characteristics we model ($T_{\rm LL}=0.9$ instead of $0.25$ for the normal line) currently does not exist. However, its implementation can in principle be possible in the future, by using an improved optical fiber or high-efficiency quantum teleportation.

When Bob sends a frame to Alice, the switches are in crossed positions (FOSb: $1\rightarrow4$ and FOSa: $2\rightarrow3$) so that the frame essentially traverses the quantum channel undisturbed. The tap is used for obtaining polarization information and synchronization, required later in preparation of the Trojan-horse pulses. Since the pulses in the forward path are relatively bright, a few photons stolen would not be noticed by Alice. 

For the return path, i.e.,\ from Alice to Bob, Eve manipulates the slots as determined by the attack pattern of figure~\ref{fstrtgy}(b). This pattern is essentially a repetition of the triad $\{N_{ab}$, $N_{el}$, $N_{ss}\}$ imposed in the reverse direction (i.e.,\ going from $N_f$ to $1$) on an entire QKD frame. The number of unbroken triads that can fit inside a frame is $k=\left\lfloor N_f/\left(N_{ab}+N_{el}+N_{ss}\right)\right\rfloor$, where $\left\lfloor \cdot \right\rfloor$ denotes the \texttt{floor} operation. This leaves exactly $N_u = N_f-k \left(N_{ab}+N_{el}+N_{ss}\right)$ \emph{unaccounted} slots in the beginning of the frame; if $N_u>N_{ab}$, then we add yet another attack burst $N_{ab}$ and extinguish the remaining $N_{el0}=N_u-N_{ab}$ slots, as also shown in figure~\ref{fstrtgy}(a) with $k=4$ and $N_u=151$. Otherwise, we simply extinguish $N_{el0}=N_u$ slots.

Using this pattern, Eve physically manipulates the frame in the following way: slots up to $N_{el0}$ are extinguished by being directed onto a beam dump (FOSa: $3\rightarrow2$ and FOSb $4\rightarrow2$). The next $N_{ab}+N_{ss}$ slots pass through the low-loss line (both FOSa and FOSb in positions $3\rightarrow1$) to Bob. Using the Trojan-horse attack apparatus (see figure~\ref{fimplmntn}), Eve reads Bob's PM settings for the \emph{attack burst}, i.e.,\ the first $N_{ab}$ of these slots. The remaining $N_{ss}$ slots, or the \emph{substitution sequence}, simply travel to Bob via the low-loss line. The switches then flip again for an \emph{extinguished length} of $N_{el}$ slots. This sequence is repeated until the end of the frame is reached with the last $N_{ab}$ gates always attacked. Attacking the last few slots causes less afterpulsing, because the detector gates are not applied after the frame end.

\end{document}